\documentclass[aip,reprint,twocolumn,superscriptaddress,showpacs,floatfix]{revtex4-1}
\usepackage{bm}
\usepackage{amssymb}
\usepackage{epsfig}
\usepackage{amsmath}
\usepackage{color}
\usepackage{graphicx}
\usepackage{epstopdf}
\usepackage{txfonts}

\begin{document}

\title{Emergence, evolution, and control of multistability in a hybrid topological quantum/classical system}

\author{Guanglei Wang}
\affiliation{School of Electrical, Computer, and Energy Engineering, Arizona
State University, Tempe, AZ 85287, USA}

\author{Hongya Xu}
\affiliation{School of Electrical, Computer, and Energy Engineering, Arizona
State University, Tempe, AZ 85287, USA}

\author{Ying-Cheng Lai} \email{Ying-Cheng.Lai@asu.edu}
\affiliation{School of Electrical, Computer, and Energy Engineering, Arizona
State University, Tempe, AZ 85287, USA}
\affiliation{Department of Physics, Arizona State University, Tempe, AZ 85287,
USA}

\date{\today}

\begin{abstract}

We present a novel class of nonlinear dynamical systems - a hybrid of 
relativistic quantum and classical systems, and demonstrate that 
multistability is ubiquitous. A representative setting is coupled systems 
of a topological insulator and an insulating ferromagnet, where the former 
possesses an insulating bulk with topologically protected, dissipationless, 
and conducting surface electronic states governed by the relativistic
quantum Dirac Hamiltonian and latter is described by the nonlinear classical 
evolution of its magnetization vector. The interactions between the two are 
essentially the spin transfer torque from the topological insulator to the 
ferromagnet and the local proximity induced exchange coupling in the opposite
direction. The hybrid system exhibits a rich variety of nonlinear dynamical 
phenomena besides multistability such as bifurcations, chaos, and phase 
synchronization. The degree of multistability can be controlled by an 
external voltage. In the case of two coexisting states, the system is 
effectively binary, opening a door to exploitation for developing spintronic 
memory devices. Because of the dissipationless and spin-momentum locking 
nature of the surface currents of the topological insulator, little power 
is needed for generating a significant current, making the system appealing 
for potential applications in next generation of low power memory devices.

\end{abstract}

\maketitle

\textbf{Topological quantum materials are a frontier area in condensed matter
physics and material sciences. A representative class of such materials is 
topological insulators, which have an insulating bulk but possess 
dissipationless conducting electronic states on the surface. For a 
three-dimensional topological insulator (3D TI) such as Bi$_2$Se$_3$, the 
surface states have a topological origin with a perfect spin-momentum locking,
effectively eliminating backscattering from non-magnetic impurities and 
generating electronic ``highways'' with highly efficient transport. 
The surface states can generally be described
by a two-dimensional Dirac Hamiltonian in relativistic quantum mechanics.
When a piece of insulating, ferromagnetic material is placed on top of a
topological insulator, two things can happen. Firstly, there is a
spin-transfer torque from the spin-polarized surface current of
the topological insulator to the ferromagnet, modulating its magnetization
and making it evolve dynamically. Secondly, the ferromagnet generates an
exchange coupling to the Hamiltonian of the topological insulator, reducing
its quantum transmission from unity and rendering it time dependent. Due to
these two distinct types of interactions in the opposite directions, the
coupled system of topological insulator and ferromagnet constitutes a novel
class of nonlinear dynamical systems - a hybrid type of systems where a
relativistic quantum description of the surface states of the 3D TI and a 
classical modeling of the ferromagnet based on the LLG 
(Landau-Lifshitz-Gilbert) equation are
necessary. The hybrid dynamical system can exhibit a rich variety of
nonlinear dynamical phenomena and has potential applications in spintronics.
Here we review some recent results in the study of such systems, focusing on
multistability. In particular, we demonstrate that multistability can emerge
in open parameter regions and is therefore ubiquitous in the hybrid
relativistic quantum/classical systems. The degree of multistability
as characterized by the ratio of the basin volumes of the multiple coexisting
states can be externally controlled (e.g., by systematically varying the
driving frequency of the external voltage). The controlled multistability
is effectively switchable binary states that can be exploited for developing
spintronic memory. For example, in spintronics, the multiple stable states
of the magnetization are essential for realizing efficient switching and
information storage (e.g., magnetoresistive random-access memory
- MRAM, with a read-out mechanism based on the giant magnetoresistance
effect). MRAM is widely considered to be the next generation of the universal
memory after the current FLASH memory devices. A key challenge of the current
MRAM technology lies in its significant power consumption necessary for
writing or changing the direction of the magnetization. The spin-transfer 
torque based magnetic tunnel junction configuration has been developed for 
power-efficient MRAMs. For this type of applications the coupled TI-ferromagnet 
system represents a paradigmatic setting where only extremely low driving 
power is needed for high performance because of the dissipationless 
nature of the spin-polarized surface currents of the TI.}

\section{Introduction} \label{sec:intro}

The demands for ever increasing processing speed and diminishing power
consumption have resulted in the conceptualization and emergence of physical 
systems that involve topological quantum states. Such a system can appear in 
a hybrid form: one component effectively obeying quantum mechanics on a 
relatively large length scale ($\sim\mu$m) while another governed
by classical or semiclassical equations of motion, with interactions 
of distinct physical origin in opposite directions. The topological states 
based quantum component acts as an electronic highway, where electrons can 
sustain a dissipationless, spin-polarized current under a small electrical 
driving field. The dynamics of the classical component can be nonlinear, 
leading to a novel class of nonlinear dynamical systems. We believe such 
systems represent a new paradigm for research on nonlinear dynamics. The 
purpose of this mini-review article is to introduce a class of coupled 
topological quantum/classical hybrid systems, and discuss the dynamical 
behaviors with a special focus on the issue of multistability. Significant 
potential applications will also be articulated. 

The practical motivations for studying the class of topological 
quantum/classical hybrid systems are the following. The tremendous advances 
of information technology relied on the development of hardwares. Before 
2003, the clock speed of CPU increases exponentially as predicted by 
Moore's law and Dennard scaling~\cite{Sutter:2005}. As we approach the 
end of Moore's law, mobile Internet rises, which requires more power 
efficient and reliable hardwares. Mobile Internet connects every user 
into the network and provides enormous data, leading to the emergence of
today's most rapidly growing technology - artificial intelligence 
(AI)~\cite{CV:2015,LBH:2015,JM:2015}. The orders of magnitude increases in 
abilities to collect, store and analyze information require new physical 
principles, designs, and methods - ``more is different''~\cite{Anderson:1972}. 
In the technological development, quantum mechanics has become increasingly 
relevant and critical. In general, when the device size approaches the 
scale of about 10nm, quantum effects become important. In the current 
mesoscopic era, a hybrid systems description 
incorporating classical and quantum effects is essential. For example, 
quantum corrections such as energy quantizations and anisotropic mass 
are necessary in the fabrication of CMOS~\cite{USK:2008}, the elementary 
building block of CPU and GPU. In terms of memory devices, the magnetic 
tunnel junction (MTJ) based spin-transfer torque random access memory 
(STTRAM)~\cite{HYYBHYYSHF:2005,TKMYHMOYIT:2010} is becoming a promising 
complement to solid state drives, whose core is a tunnel junction design 
with multilayer stacks, where a layer of ferromagnetic materials with a 
fixed magnetization is used to polarize the spin of the injected current. 
As a fully quantum phenomenon, the spin polarized current tunnels through 
a barrier to drive the magnetization in the magnetically soft ferromagnetic 
layer - the so-called free layer for information storage. 

The basic principles of STTRAM was proposed about two decades ago in
the context of spintronics before the discovery of topological insulators 
(TIs)~\cite{BHZ:2006,HK:2010,QZ:2011}. In addition, a current-induced 
spin-orbit torque mechanism~\cite{GM:2011} was proposed as an alternative 
way to harness the magnetization of conducting magnetic or magnetically 
doped materials with large spin-orbit coupling. With the advance of 
3D TIs, much more efficient operations are anticipated with giant  
spin-orbit torque and spin-transfer torque. Particularly, in comparison 
with the conventional spin-orbit torque settings in heavy metals with a 
strong Rashba type of spin-orbit 
coupling~\cite{MGARSPVG:2010,MGGZCABRSG:2011,PWBLCKS:2010,WM:2012}, the 
spin density in 3D topological insulator based systems can be enhanced 
substantially by the factor $\hbar v_F/\alpha_R\gg1$, where $v_F$ is the 
Fermi velocity in the TI and $\alpha_R$ is the strength of the Rashba 
spin-orbit coupling in two-dimensional electron gas (2DEG) 
systems~\cite{CMSSN:2015}. The two-layer stack configurations of 
TI/ferromagnet~\cite{DLSK:2015,TST:2015,JLJMLZNMSW:2015} or 
TI/anti-ferromagnet~\cite{RNS:2017,LDSK:2017} have recently been articulated,
which allow for the magnetization of the depositing magnetic materials to
be controlled and the transport of spin-polarized states on the surface 
of the 3D TIs to be modulated.

\begin{figure*}
\centering
\includegraphics[width=\linewidth]{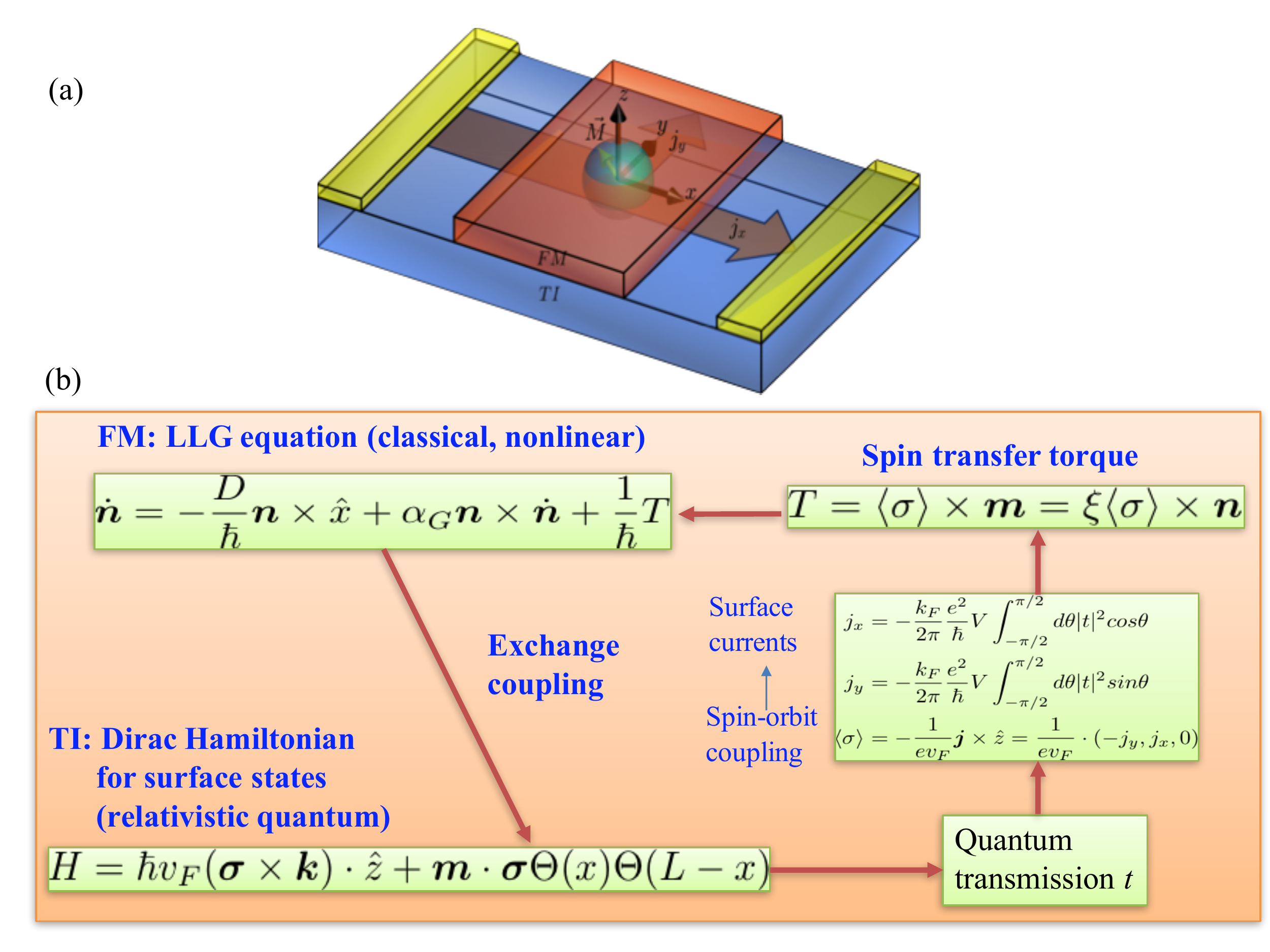}
\caption{ {\bf Schematic illustration of a representative relativistic 
quantum/classical hybrid system, the physical interactions, and the 
dynamical evolution of the states}. (a) A coupled ferromagnet and topological 
insulator (TI) system, where the ferromagnet is deposited on the top
of the TI. (b) The physical interactions: spin-transfer torque (from TI
to ferromagnet) and exchange coupling (from ferromagnet to TI). The dynamical 
evolution of the ferromagnet is described by the classical, nonlinear LLG 
equation, and the dissipationless, spin-polarized surface states of the 
TI are determined by the Dirac Hamiltonian. Refer to Secs.~\ref{sec:TI} 
and \ref{sec:STT} for the meanings of the mathematical notations and 
equations.}   
\label{fig:FM_TI_Schematic}
\end{figure*}

To be concrete, in this paper we consider the ferromagnet-TI configuration 
and focus on the dynamics of the magnetization in the insulating ferromagnet 
and the two orthogonal current components on the surface of the TI: one 
along and another perpendicular to the direction of an externally applied
electrical field. A schematic illustration of the system is presented in 
Fig.~\ref{fig:FM_TI_Schematic}(a), where a rectangular shape of the ferromagnet
is deposited on the top of a TI. For the ferromagnet, the dynamical variable 
is the magnetization vector ${\bf M}$, whose evolution is governed by the 
classical Landau-Lifshitz-Gilbert (LLG) equation~\cite{Slonczewski:1996}, 
which is nonlinear. The TI, as will be described in Sec.~\ref{sec:TI}, 
hosts massless spin-$1/2$ quasiparticles in the low-energy regime ($\sim$meV) 
on its surface and generates spin polarized surface currents when a weak 
electrical field is applied~\cite{BHZ:2006,HK:2010,QZ:2011}. In fact, the 
surface states of the TI are described by the Dirac Hamiltonian, rendering 
it effectively relativistic quantum. Physically, the interactions between 
TI and ferromagnet can be described, as follows. The robust spin polarized 
current on the surface of the TI generates~\cite{MLRGMFVMKS:2014} a 
strong spin-transfer torque~\cite{RS:2008} to the ferromagnet, inducing 
a change in its magnetization vector. The ferromagnet, in turn, generates 
a proximity induced exchange field in the TI. As a result, there is an 
exchange coupling term in the surface Dirac Hamiltonian of the TI, which 
modulates the quantum transmission and leads to a change in the surface 
current. The two-way interactions between the ferromagnet and TI are 
schematically illustrated in Fig.~\ref{fig:FM_TI_Schematic}(b), where the 
whole coupled system is of a hybrid type: effectively relativistic quantum 
TI and classical ferromagnet. The interactions render time dependent the 
dynamical variables in both TI and ferromagnet: the surface current for 
the former and the magnetization vector for the latter. Due to the 
intrinsic and externally spin-transfer torque induced nonlinearity of
the LLG equation, the whole configuration represents a nonlinear dynamical
system, in which a rich variety of phenomena such as bifurcations,
chaos, synchronization, and multistability can arise~\cite{WXL:2016}. 

Following the theme of the Focus Issue, in this paper we focus on the 
emergence, evolution and control of multistability. When a small external 
voltage with both a dc and ac component is applied to the TI, a robust 
spin-polarized surface current rises, as shown in 
Fig.~\ref{fig:FM_TI_Schematic}(a). In certain open parameter regions, the 
magnetization can exhibit two coexisting stable states (attractors) with 
distinct magnetic orientations, each having a basin of attraction. As the 
phase space for the magnetization is a spherical surface, the basin areas 
of the two attractors are well defined. (This is different from typical 
dissipative nonlinear dynamical systems in which the basin of attraction of 
an attractor has an infinite phase space volume~\cite{Ott:book}.) 
As an external parameter, e.g., the frequency of the ac voltage, is 
varied, the relative basin areas of the two attractors can be continuously
modulated. In fact, as the parameter is changed systematically, both 
birth and death of multistability can be demonstrated, rendering feasible
manipulation or control of multistability. While some of these results 
have appeared recently~\cite{WXL:2016}, here we focus on those that have
not been published. 

In Sec.~\ref{sec:TI}, we provide a concise introduction to the basics of 
TIs and the Dirac Hamiltonian with a focus on the physical pictures of 
the emergence of strong, spin-polarized surface states. In Sec.~\ref{sec:STT},
we describe the mechanism of spin transfer torque and the rules of the dynamical
evolution of the TI-ferromagnet coupled system in terms of the LLG equation 
and the quantum transmission of the TI. In Sec.~\ref{sec:multistability}, we 
present results of multistability, followed by a discussion of potential 
applications in Sec.~\ref{sec:discussion}.    

\section{Topological insulators} \label{sec:TI}

\begin{figure}
\centering
\includegraphics[width=\linewidth]{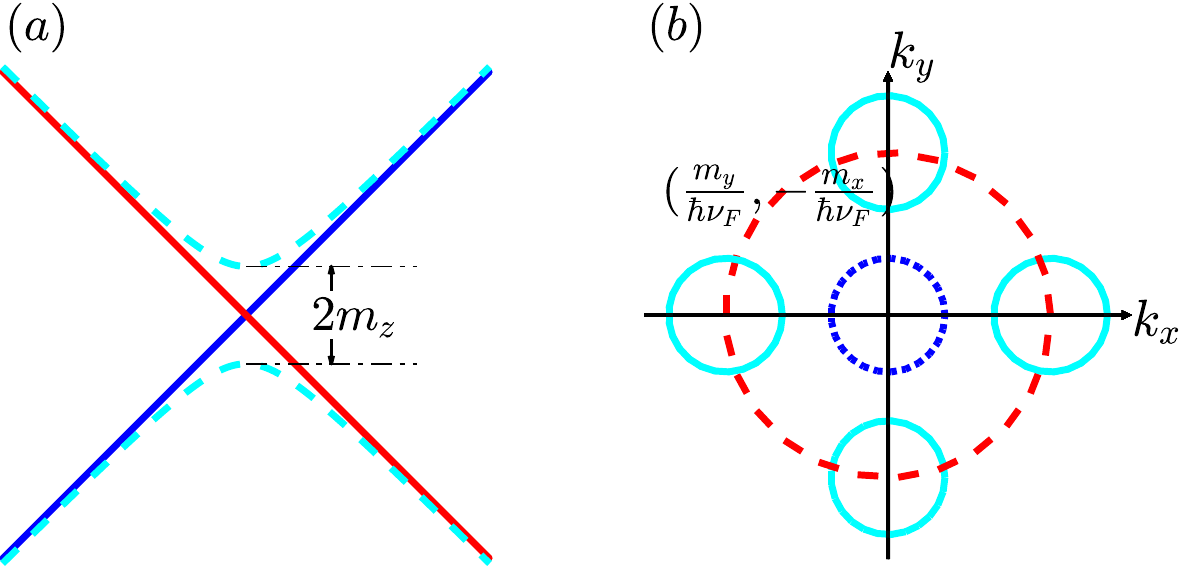}
\caption{{\bf Surface states of topological insulators and the effect of
magnetization}. (a) Red and blue lines represent the linear dispersion
relation of the surface states of an ideal TI without a gap opening 
perturbation, with the slope $\hbar\nu_F$. The two states have opposite 
spin polarizations. The presence of a perpendicular magnetization vector 
$m_z$ leads to a gap opening of size proportional to $2m_z$, making the 
dispersion relation hyperbolic. (b) Magnetization vectors in the plane of 
the TI, i.e., $m_x$ and $m_y$, will shift the central position of the Fermi 
surface in the wave vector space, inducing an asymmetry with respect to 
either $k_y=0$ or $k_x=0$.}
\label{fig:TI}
\end{figure}

One of the most remarkable breakthroughs in condensed matter physics in the 
last decade is the theoretical prediction~\cite{BHZ:2006,FK:2007,ZLQDFZ:2009} 
and the subsequent experimental realization~\cite{KWBRBMQZ:2007,HQWXHCH:2008,
XQHWPLBGHC:2009} of TIs~\cite{Moore:2010,HK:2010,QZ:2011}. TIs are one 
emergent phase of the material that has a bulk band gap so its interior is 
an insulator but with gapless surface states within the bulk band gap. The 
surface states are protected by the time-reversal symmetry and therefore are 
robust against backscattering from impurities, which are practically appealing
to developing dissipationless or low-power electronics. Moreover, the surface 
states possess a perfect spin-momentum locking, in which the spin orientation 
and the direction of the momentum is invariant during the propagation. 

TIs are representative of topologically protected phases of 
matter~\cite{TKNN:1982,Haldane:1988}, one theme of last year's Nobel 
prize~\cite{NP:2016}. The prediction and experimental realization of TIs
benefited from the well known quantum Hall effect~\cite{KDP:1980}, also
a topological quantum order. The topological phases of matter not only are
of fundamental importance, but also have potential applications in 
electronics and spintronics~\cite{PM:2012}. According to the bulk-edge 
correspondence in topological field theory, gapless edge states exist at the 
boundary between two materials with different bulk, topologically invariant 
numbers~\cite{HK:2010}. The edge states are protected by the topological 
properties of the bulk band structures, and thus are extremely robust 
against local perturbations. 
Depending on the detailed system setting, there are remarkable properties 
associated with the edge states such as perfect conductance, uni-directional 
transportation, and spin-momentum coupling~\cite{BHZ:2006,HK:2010,QZ:2011}.

We learned from elementary physics that a perpendicular magnetic field applied 
to a conductor subject to a longitudinal electrical field will induce a 
transverse voltage - the classic Hall effect~\cite{Hall:1879}. In 1980, it 
was discovered that, for a 2DEG at low temperatures, under a strong magnetic 
field the Hall conductivity is quantized exactly at the integer multiple
of the fundamental conductivity $e^2/h$, where $-e$ is the electronic 
charge and $h$ is the Planck constant. This is the integer quantum Hall
effect (usually referred to as QHE)~\cite{KDP:1980}. Different from the 
classical Hall effect, the quantized Hall conductivity is fundamentally 
a quantum phenomenon occurring at the macroscopic scale~\cite{Laughlin:1981}. 
Soon after, new states of matter such as the fractional quantum Hall effect 
(FQHE)~\cite{TSG:1982,Laughlin:1983,Nobel:1998,STG:1999}, quantum anomalous 
Hall effect (QAHE)~\cite{Haldane:1988}, and the quantum spin Hall effect 
(QSHE)~\cite{Hirsch:1999,MNZ:2003,SCNSJM:2004,BZ:2006} were discovered. 
Besides their fundamental significance, such discoveries have led to an
unprecedented way to understand and explore the phases of matter through a 
connection of two seemingly unrelated fields: condensed matter physics 
and topology.

Heuristically, QHE can be understood in terms of the Landau levels - 
energy levels formed due to a strong magnetic field. Classically, an
electron will precess under such a field. Quantum mechanically, only
the orbits whose circumference is an integer multiple of $2\pi$ can emerge 
(constructive interference), leading to the Landau levels. QHE can be
understood as a result of the Fermi level's crossing through various
Landau levels~\cite{Datta:book} as, e.g., the strength of the external
magnetic field is increased.     

At a deeper level, the robustness of the quantized conductivity in QHE 
can be understood as a topological effect. To understand the concept of
topology in physical systems, we consider a two-dimensional closed surface 
characterized by a Gaussian curvature. According to the Gauss-Bonnet 
theorem, the number of holes associated with a compact surface 
without a boundary can be written as a closed surface integral:
\begin{displaymath} 
4\pi(1-g) = \oiint (\mbox{Gaussian curvature})\cdot da,
\end{displaymath} 
where we have $g = 0$ for a spherical surface and $g = 1$ for a donut 
surface (a two-dimensional torus). Different types of geometrical surfaces 
can thus be characterized by a single number. This idea has been extended
to physics with a proper definition of geometry in terms of the quantum 
eigenstates, where a generalized Gauss-Bonnet formula by Chern applies. 
In particular, consider the band structure of a 2D conductor or 
a 2DEG. Let $|u_m({\bf k})\rangle$ be the Bloch wavefunction associated 
with the $m$th band. The underlying Berry connection is 
\begin{equation} \label{eq:Berry_curvature}
{\bf A}_m = i \langle u_m({\bf k})|\nabla_{\bf k}|u_m({\bf k})\rangle.
\end{equation}
The Berry phase is given by
\begin{equation} \label{eq:Berry_phase}
\Phi_B = \oint {\bf A}_m \cdot d{\bf l} = 
\oiint (\nabla\times {\bf A}_m)\cdot {\bf ds},
\end{equation}
where ${\bf F_m} = \nabla\times {\bf A}_m$ is the Berry curvature.
The total Berry flux associated with the $m$th band in the Brillouin 
zone is 
\begin{equation}
n_m = \frac{1}{2\pi} \iint d^2{\bf k}\cdot {\bf F_m}, 
\end{equation}
which is the Chern number~\cite{TKNN:1982,HK:2010}. Let $N$ be the number 
of occupied bands. The total Chern number is 
\begin{equation} \label{eq:Chern_number}
n = \sum^{N}_{m=1} n_m.
\end{equation}
It was proved~\cite{TKNN:1982} that the Hall conductivity is given by
\begin{equation} \label{eq:Hall_conductivity}
\sigma_{xy} = n \frac{e^2}{h}.
\end{equation}
As the magnetic field strength is increased, the integer Chern number 
increases, one at a time, leading to a series of plateaus in the 
conductivity plot. The Chern number is a topological invariant: it cannot 
change when the underlying Hamiltonian varies smoothly~\cite{TKNN:1982}.
This leads to robust quantization of the Hall conductance. Due to its 
topological nature and the time-reversal symmetry breaking, dissipationless 
chiral edge conducting channels emerge at the interface between the integer 
quantum Hall state and vacuum, which appears to be promising for developing
low-power electronics but with the requirement of a strong external magnetic 
field. Nevertheless, the topological ideas developed in the context of QHE 
turned out to have a far-reaching impact on pursuing distinct topological 
phases of matter.

The QSHE represents a preliminary manifestation of TIs with a time reversal 
symmetry, as a 2D TI, essentially a quantum spin Hall state, was predicted in 
the CdTe-HgTe-CdTe quantum well system~\cite{BHZ:2006} and experimentally 
realized~\cite{KWBRBMQZ:2007}. Both CdTe and HgTe have a zinc blende crystal 
structure and have minimum band gaps about the $\Gamma$ point. The origins 
of the conduction and valance bands are $s$ and $p$ atomic orbitals. 
Compared with CdTe, HgTe can support an inverted band structure about the 
$\Gamma$ point, i.e., CdTe has an $s$-type conduction band and a $p$-type 
valance band while HgTe has a $p$-type conduction band and an $s$-type 
valance band. Mathematically, the inversion is equivalent to a change in 
the sign of the effective mass. The difference is in fact a result of  
the strong spin-orbit coupling in HgTe, which can reduce the band gap and 
even invert the bandstructure through orbital splitting. (In experiments, 
the band gap can be widened by increasing the size of the quantum confinement.)
Intuitively, a CdTe-HgTe-CdTe quantum well can be thought of as making a
replacement of the layers of Cd atoms by Hg. When the HgTe layer is thin, 
the properties of CdTe is dominant and the quantum well is in the normal 
regime. As one increases the thickness of the HgTe layer, eventually the 
configuration of the conduction and valance bands will become similar
to that of HgTe so the quantum well is in the inverted regime. The interface 
joining the two materials with the inverted configuration acts as a domain 
wall and can potentially harbor novel electronic states described by a 
distinct Hamiltonian. 

In general, regardless of the material parameters, the Hamiltonian 
can be written in the momentum space as $H=H(k_x,k_y,-i\partial_z)$, 
where $z$ stands for the stacking direction so $k_z$ is not a good quantum 
number. Integrating this Hamiltonian within several dominant $z$ base 
states, one can arrive at an effective Hamiltonian for the 2D TI, which 
is characterized by $k_x$ and $k_y$. Edge states can be obtained 
by imposing periodic boundary conditions along one direction and open 
boundary conditions in another. The inverted band structure, 
i.e., the basis, and the time-reversal symmetry guarantee the gaplessness 
of the edge state while the detailed form of the open boundary plays 
a secondary role only. For a pedagogical review of 2D TIs, 
see Ref.~\cite{KBWHLQZ:2008}.

Parallel to the study of 2D TIs, there were efforts in uncovering 3D TIs,
for which Bi$_{1-x}$Sb$_x$ was theoretically proposed to be a 
candidate~\cite{FK:2007,MNZ:2003}. The prediction was confirmed 
experimentally shortly after~\cite{HQWXHCH:2008}. However, since 
Bi$_{1-x}$Sb$_x$ is an alloy with random substitutional disorders, its
surface states are quite complicated, rendering difficult a description
based on an effective model. The attention was then turned to finding 
3D TIs in stoichiometric crystals with simple surface states, leading 
to the discovery~\cite{XQHWPLBGHC:2009,ZLQDFZ:2009} of Bi$_2$Se$_3$.
In particular, it was experimentally observed~\cite{XQHWPLBGHC:2009}
that there is a single Dirac cone on the surface of Bi$_2$Se$_3$. 
A low-energy effective model was then proposed~\cite{ZLQDFZ:2009} for 
Bi$_2$Se$_3$ as a 3D TI, in which spin-orbit coupling was identified as
the mechanism to invert the bands in Bi$_2$Se$_3$ and the four most 
relevant bands about the $\Gamma$ point were used to construct an 
effective bulk Hamiltonian. A generic form of the Hamiltonian was written
down in the space constituting the four bases up to the order of
$O(\bm{k}^2)$, constrained by a number of symmetries: 
the time-reversal, the inversion, and the three-fold rotational symmetries. 
The parameter associated with each term was determined by fitting the 
dispersion relation with the {\it ab initio} computational results. The 
surface states can be obtained by imposing constraints along one direction, 
which mathematically entails replacing $k_z$ by $-i\partial_z$ while 
keeping the states along the other two directions oscillatory. The 
effective surface Hamiltonian can be calculated by projecting the bulk 
Hamiltonian onto the surface states. At the present, Bi$_2$Se$_3$
is one of the most commonly studied 3D TIs, which possesses gapless Dirac 
surface states protected by the time-reversal symmetry and a bulk band 
gap up to $0.3\mbox{eV}$ (equivalent to $3000\mbox{K}$ - far higher 
than the room temperature).

The widely used Hamiltonian for the surface states of an ideal 3D TI is
\begin{equation} \label{eq:H}
H=\hbar v_F(\bm{\sigma}\times\bm{k})\cdot\hat{z},
\end{equation}
where $v_F$ is the Fermi velocity of the surface states 
($v_F\approx6.2 \times 10^5\mbox{m}/\mbox{s}$ for Bi$_2$Se$_3$), and 
$\bm{\sigma}=(\sigma_x,\sigma_y,\sigma_z)$ are the Pauli matrices describing 
the spin of the surface electron~\cite{ZLQDFZ:2009}. An elementary 
calculation shows that the dispersion relation of this surface Hamiltonian 
indeed has the structure of a Dirac cone. In proximity to a ferromagnet with 
a magnetization $\bm{m}$, an extra term $\bm{m}\cdot\bm{\sigma}$ in the 
Hamiltonian (\ref{eq:H}) will be induced. Due to the breaking of the time 
reversal symmetry by the exchange field, gap opening for the surface states 
will occur and backward scatterings will no longer be forbidden. The 
proximity induced exchange field will thus modulate the charge transport 
behaviors of the surface states and the underlying spin density by disturbing 
the spin texture, which in turn can be a driving source of the nonlinear 
dynamic magnetization in the adjacent ferromagnetic cap layer through a 
spin-transfer torque. 

\section{Spin-transfer torque} \label{sec:STT}

\begin{figure}
\includegraphics[width=\linewidth]{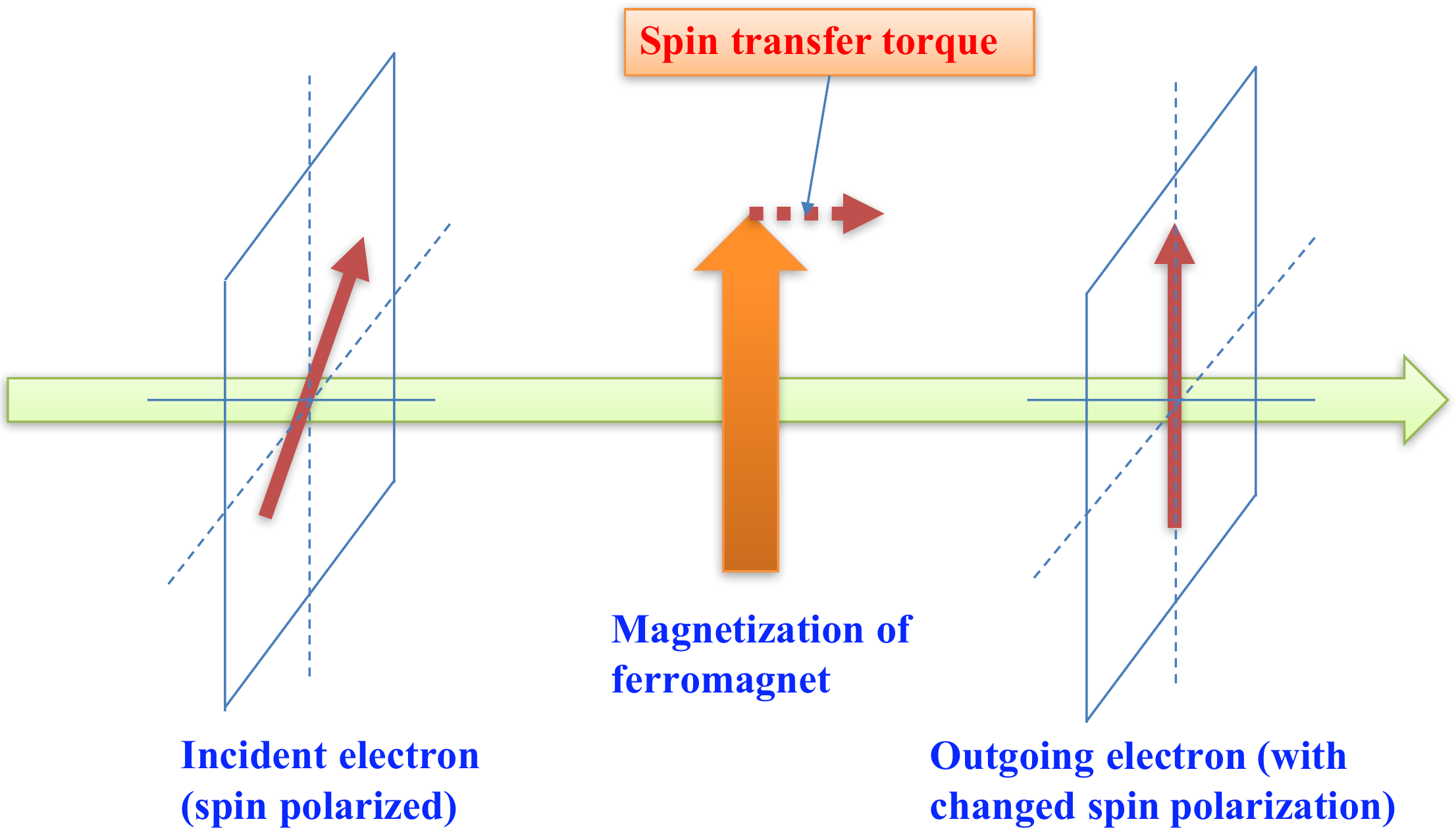}
\caption{ {\bf Schematic illustration of spin-transfer torque}. The
large green arrow indicates an electron flow under an external longitudinal
voltage. The electrons are spin polarized (e.g., the surface states of 
a 3D TI). As the electrons pass through a region in which a magnetization
vector is present (e.g., a ferromagnetic region), it exerts a torque on
the magnetization, due to which the spin polarization of the outgoing 
electrons is changed, henceforth the term ``spin-transfer torque.''}   
\label{fig:STT}
\end{figure}

Spin-transfer torque~\cite{RS:2008} is a major subject of research in 
spintronics~\cite{LCG:2014}, a field aiming to understand and exploit 
the spin degree of freedom of electrons beyond the conventional charge 
degree of freedom. Intuitively, spin-transfer torque is nothing but the  
exchange interaction between two magnetization vectors. In particular, when 
two magnetization vectors are brought close to each other, they tend to 
align or anti-align with each other to evolve into a lower energy state. In our
coupled TI-ferromagnet system, one magnetization vector is the net contribution
of the spin-polarized current on the surface of the TI, and the other 
comes from the ferromagnet. A setting to generate a spin-transfer torque 
is schematically illustrated in Fig.~\ref{fig:STT}. The basic physical 
picture can be described, as follows. When a normal current flows near or 
within a region in which a strong magnetization is present, the spin 
associated with the current will be partially polarized. This implies
that, by the law of action and reaction, a spin-polarized current will 
exert an torque on the magnetization - the spin-transfer torque. Such a 
torque will induce oscillations, inversion and other dynamical behaviors 
of the magnetization. The ability to manipulate magnetization is 
critical to applications, especially in developing memory devices. 

\begin{figure}
\includegraphics[width=\linewidth]{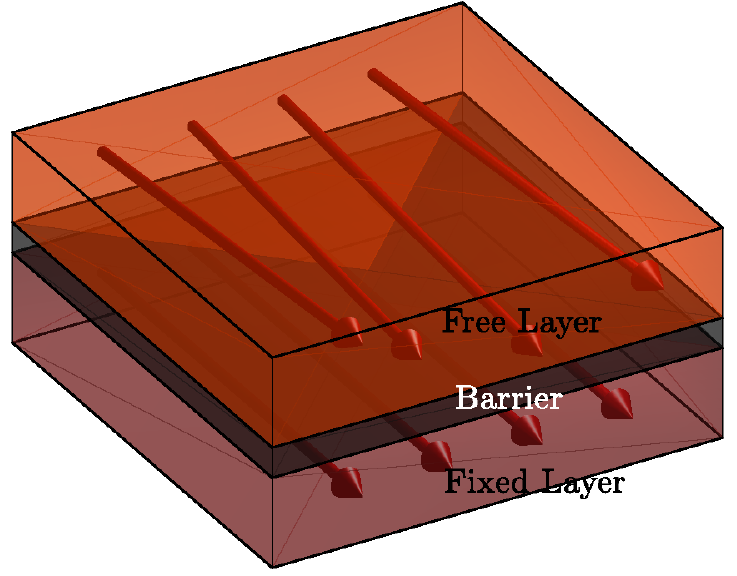}
\caption{ {\bf Schematic illustration of a magnetic tunnel junction (MTJ)}.
A typical MTJ consists of a fixed layer, a barrier and a free layer. 
An electrical current is injected vertically into the fixed layer and 
comes out from the free layer. The average spin of the current is 
polarized by the fixed layer, which tunnels through the insulating barrier 
and drives the magnetization of the free layer. This configuration can be 
used to modify and detect the magnetization direction in the free layer 
through the giant magnetoresistance (GMR) effect.}
\label{fig:MTJ}
\end{figure}

Conventionally, a three-layer magnetic tunnel junction (MTJ) is used to
study spin-transfer torque~\cite{GNSTOAGHBZ:2012,BKO:2012,LCG:2014}, in 
which the fixed layer is a ferromagnet with a permanent magnetization.
A schematic illustration of an MTJ is presented in Fig.~\ref{fig:MTJ}. 
When a normal current is injected into this layer, the spin will be 
partially polarized due to the individual exchange interaction between 
each single electron spin and the magnetization. As a result, there is 
randomness in the polarization with no definite correlation 
between the direction of the electron spin and momentum. To separate the 
fixed from the free layer, a thin insulating separation barrier is 
needed to form a tunnel junction. The current will travel through the 
insulating barrier via the mechanism of quantum tunneling, during which
the direction of spin will not be altered insofar as the insulating 
barrier does not have any magnetic impurity. After the tunneling, the 
spin-polarized current will exert a spin-transfer torque on the 
magnetization in the free ferromagnetic layer, modifying the information
stored. Readout of the information, i.e., the direction of the
magnetization, can be realized by exploiting the giant magnetoresistance 
effect~\cite{BBFVPECFC:1988,BGSZ:1989,Nobel:2007}. 

An issue with MTJ is that a very large current is needed to reorient 
the magnetization, motivating efforts to explore alternative mechanisms 
with a lower energy requirement. One mechanism was discovered in 
ferromagnet/heavy-metal bilayer heterostructures~\cite{MGGZC:2011,LPLTRB:2012}.
The strong Rashba spin-orbit coupling in many heavy metals can be exploited
through the mechanism of spin-orbit torque to generate spin-polarized 
currents via the Edelstein effect, which are generally much stronger than 
the exchange interaction between the fixed layer and current spin in the 
conventional MTJ. Moreover, the current in the configuration needs no longer
to be restricted to the perpendicular direction, but can have any 
orientation within the film plane. This new geometrical degree of freedom 
enables unconventional strategies for manipulating magnetization. For
example, it was discovered~\cite{SJLBAPBMG:2016} that the domain wall motions
(essentially the dynamics of magnetization) in the covering magnetic free layer 
have a sensitive dependence on the spatial distribution of the current 
generated spin-orbit torque.

A disadvantage of the spin-orbit torque configuration is that the heavy
metals usually suffer from substantial scatterings and the transportation
mechanisms are complex. In addition, for heavy metals, spin-orbit coupling
is essentially a higher-order effect. As a result, the currents are not 
perfectly polarized. These difficulties can be overcome by exploiting
TIs as a replacement for heavy metals. 

Mathematically, the Hamiltonian term describing the Rashba spin-orbit 
coupling has the same form as the effective surface Hamiltonian of a 3D 
TI, i.e., $\sim\bm{\sigma}\times\bm{k}$. This is basically the whole 
Hamiltonian for the surface states of 3D TI under the low energy 
approximation. The presence of an exchange field from the ferromagnetic 
cap layer will contribute a Zeeman term to the Hamiltonian. To see this 
explicitly, we consider the surface Hamiltonian of a 3D TI in the presence 
of a magnetization
\begin{equation}
\begin{aligned}
& \hbar v_F(\bm{\sigma}\times\bm{k})\cdot\hat{z}+\bm{m}\cdot\bm{\sigma}
\nonumber \\
= & \hbar v_F[\sigma_x\cdot(k_y+\frac{m_x}{\hbar v_F}) - \sigma_y \cdot
(k_x-\frac{m_y}{\hbar v_F})] + m_z\sigma_z,
\end{aligned}
\end{equation}
where the $m_z$ term only induces a band gap due to the breaking of the
time-reversal symmetry for the surface states. The $m_x$ and $m_y$ terms are 
equivalent to a shift in the center of the Fermi surface, as shown in 
Fig.~\ref{fig:TI}(b). Consequently, the surface currents and the associated 
spin densities flowing through the magnetization region will be modulated. 
In addition, as the TI contains an insulating bulk with a band gap much 
larger than the room temperature thermal fluctuations, the only conducting 
modes are those associated with the spin-polarized surface states. The 
coupled TI/ferromagnet configuration has been experimentally 
realized~\cite{MLRGMFVMKS:2014}. So far the system provides the strongest 
spin-transfer torque source - two to three orders of magnitude higher than 
that in heavy metals.

The magnetization dynamics of the ferromagnet deposited on the 3D TI
can be described by the classical LLG equation~\cite{Slonczewski:1996},
which captures the essential physical processes such as 
procession and damping of the magnetization subject to external torques.
The LLG equation is
\begin{equation} \label{eq:LLG}
\dot{\bm{n}}=-\frac D\hbar \bm{n}\times\hat{x}+\alpha_G\bm{n}\times
\dot{\bm{n}}+\frac 1\hbar \bm{T},
\end{equation}
where the first term represents the procession along the easy axis 
$\hat{x}$ and $D$ is the anisotropic energy of the ferromagnet. The 
second term describes the Gilbert damping of strength $\alpha_G$. The 
last term is the contribution from the external torque, which is the 
spin-transfer torque:
\begin{equation} \label{eq:STT}
\bm{T}=\langle\bm{\sigma}\rangle\times\bm{m}=\xi\langle\bm{\sigma}\rangle
\times\bm{n},
\end{equation}
where, for convenience, we use $\xi=|\bm{m}|$ (the magnitude of the 
magnetization) as a normalizing factor so that $\bm{n}=\bm{m}/\xi$ 
becomes a unit vector.

\section{Emergence, evolution and control of multistability} 
\label{sec:multistability}

\begin{figure*}
\centering
\includegraphics[width=0.8\linewidth]{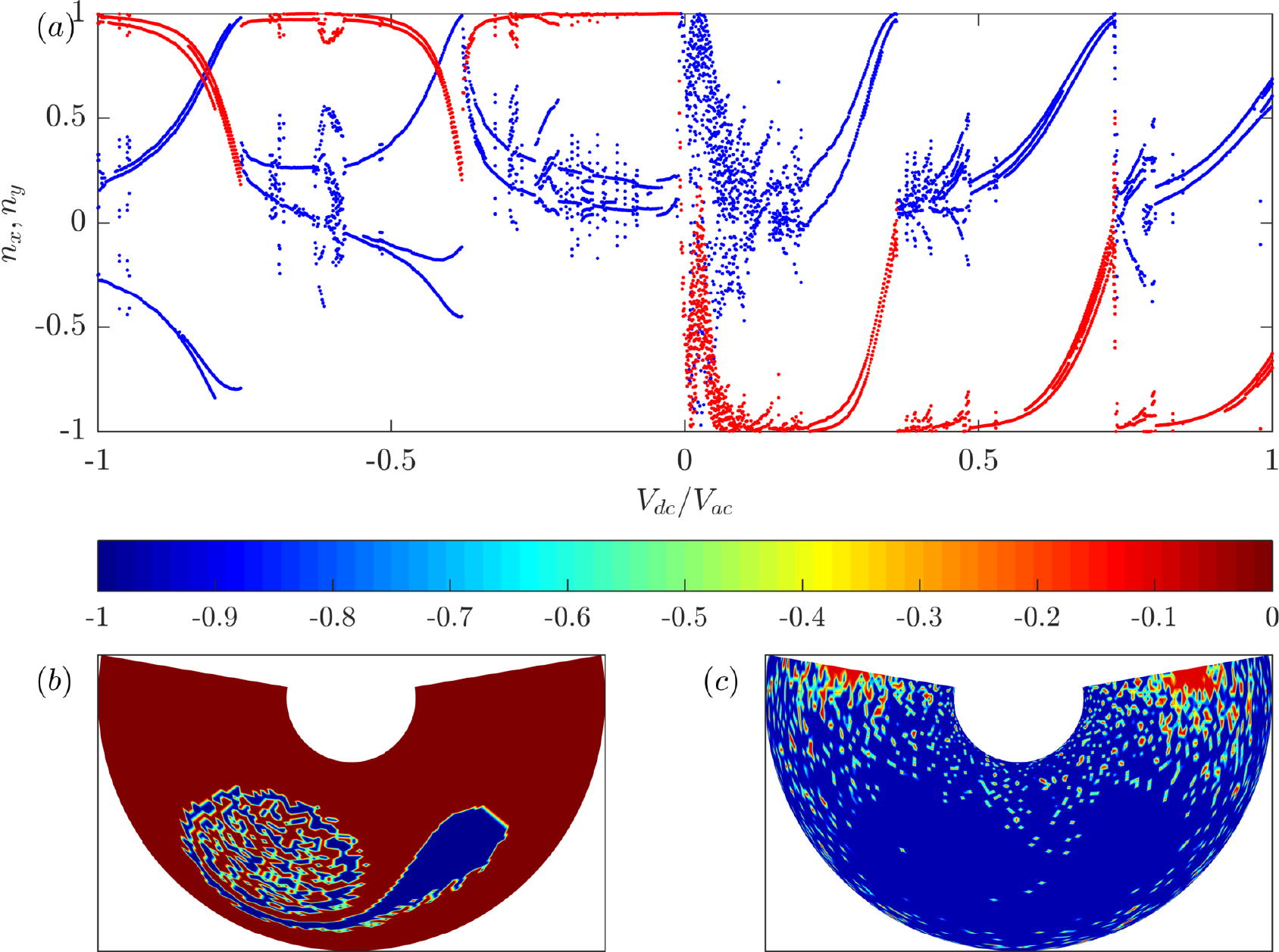}
\caption{ {\bf Bifurcations and multistability in the coupled TI-ferromagnet 
system}. (a) A bifurcation diagram for $\Omega/\omega_F=7.0$, where the
parameter ratio $V_{dc}/V_{ac}$ is swept from $-1.0$ to $1.0$. The 
system exhibits rich dynamics. In several parameter regimes, there 
are abrupt changes in the  final states. A systematic computation
with different initial conditions reveals multistability. (b-c) 
For $\Omega/\omega_F = 10.0$, typical
examples of multistability. The phase space of the normalized 
magnetization ${\bf n}$ of the ferromagnet is the surface of a unit sphere.
All possible initial conditions from the spherical surface are distinguished 
by different colors. The Albers equal-area conic projection is used to 
map the initial conditions from the spherical surface to a plane while 
preserving the area of each final state. The standard parallels of the
Albers projection are $\frac29\pi, \frac7{18}\pi$. Shown are the basins 
of two final states.}
\label{fig:bif}
\end{figure*}

To study the nonlinear dynamics of the coupled TI-ferromagnet system, we 
couple the transportation of current on the surface of TI with the oscillatory 
dynamics of the magnetization of the ferromagnet through the mechanisms of 
spin-transfer torque and exchange coupling. The various directions are defined
in Fig.~\ref{fig:FM_TI_Schematic}(a). The covering ferromagnetic material 
is insulating so that the conducting current is limited to the 2D surface 
of the TI: $\bm{j}=(j_x,j_y,0)$. In addition, the width of the device in
the $y$ direction is assumed to be large, while the system size in the 
$x$ direction is on the order of the coherent length so that the transport 
of the surface current can be described as a scattering process under a 
square magnetic potential. The typical time scale of the evolution of the 
magnetization is nanoseconds, which is much slower compared with the 
relaxation time of the surface current of the TI. We can then use the 
adiabatic approximation when modeling the dynamics of the surface electrons.
Specifically, we solve the time-independent Dirac equation with a constant 
exchange coupling term at a given time and obtain the transmission 
coefficient of surface electrons~\cite{Yokoyama:2011,SDK:2014}. The 
low-energy effective surface state Hamiltonian of the TI under a square 
magnetic potential is given by
\begin{equation} \label{eq:Hm}
H=\hbar v_F(\bm{\sigma}\times\bm{k})\cdot\hat{z}+\bm{m}\cdot\bm{\sigma}
\Theta(x)\Theta(L-x).
\end{equation}
Compared with Eq.~(\ref{eq:H}), the induced exchange field 
is modeled by a step function $\Theta(x)$ in the space to ensure that it only appears 
within the ferromagnetic region of length $L$.

To calculate the transmission coefficient through the ferromagnetic region, 
we consider the wavefunctions before entering, inside and after exiting 
the ferromagnetic region, and apply the boundary conditions at the 
interfaces of the three regions. The result is~\cite{Yokoyama:2011}
\begin{equation} \label{eq:t}
t=\frac{-4\hbar\nu_F\tilde{k}_x\cos{\theta}}{\alpha(A+ie^{i\theta}B)},
\end{equation}
where
\begin{equation*}
\begin{aligned}
A=&\alpha_2[ie^{-i\theta}\hbar\nu_F(\tilde{k}_y+i\tilde{k}_x)-E-m_z] \\
&-\alpha_1[ie^{-i\theta}\hbar\nu_F(\tilde{k}_y-i\tilde{k}_x)-E-m_z], \\
B=&\alpha_2[ie^{-i\theta}(E-m_z)-\hbar\nu_F(\tilde{k}_y-i\tilde{k}_x)]\\
&-\alpha_1[ie^{-i\theta}(E-m_z)-\hbar\nu_F(\tilde{k}_y+i\tilde{k}_x)],
\end{aligned}
\end{equation*}
and 
\begin{eqnarray}
\nonumber
E & = & \hbar\nu_Fk_F, \\ \nonumber
k_x & = & k_F\cos{\theta}, \\ \nonumber 
k_y & = & k_F\sin{\theta}, \\ \nonumber
\hbar\nu_F\tilde{k}_x & = & \sqrt{E^2-m_z^2-(\hbar\nu\tilde{k}_y)^2}, 
\\ \nonumber
\hbar\nu_F\tilde{k}_y & = & \hbar\nu_Fk_y+m_x, \\ \nonumber
\alpha & = & e^{ik_FLcos\theta}, \\ \nonumber
\alpha_1 & = & e^{i(\tilde{k}_x+m_y)L}, \\ \nonumber
\alpha_2 & = & e^{i(-\tilde{k}_x+m_y)L}. 
\end{eqnarray}
In these expressions, $E$ and $k_F$ are the Fermi energy and Fermi wave 
vector of electrons outside the ferromagnetic region, i.e., the linear 
dispersion region, and $\theta$ is the incident angle of the electron to
the ferromagnetic region. Integrating the transmission coefficient over 
the incident angle, we get the current densities through the ferromagnetic
region along the $x$ and $y$ directions as
\begin{equation} \label{eq:J}
\begin{aligned}
j_x=&-\frac{k_F}{2\pi}\frac{e^2}\hbar V\int_{-\frac\pi 2}^{\frac\pi 2}
d\theta|t|^2\cos{\theta}, \\
j_y=&-\frac{k_F}{2\pi}\frac{e^2}\hbar V\int_{-\frac\pi 2}^{\frac\pi 2}
d\theta|t|^2\sin{\theta},
\end{aligned}
\end{equation}
from which the spin density of the electrons can be calculated as
\begin{equation} \label{eq:sigma}
\langle\bm{\sigma}\rangle=-\frac1{e\nu_F}\bm{j}\times\hat{z}=\frac1{e\nu_F}
\cdot(-j_y,j_x,0)=\frac{j_x}{e\nu_F}\cdot\bm{\eta},
\end{equation}
where $V=V_{dc}+V_{ac}\cos{(\Omega t)}$ is the driving voltage along the $x$
direction and $\bm{\eta}=(\eta_x,1,0)$. A surprising result is that, the
voltage along the $x$ direction will induce a current along $y$ direction, 
which is a signature of an anomalous Hall effect~\cite{WXL:2016}. To
understand the origin of this current deviation, we examine the integrand 
of Eq.~(\ref{eq:J}) in the absence of the ferromagnet:
\begin{equation}
|t|^2=\frac{\cos^4{\theta}}{\cos^4{\theta}\cos^2{(k_xL)}
+(\sin^2{\theta}-1)^2\sin^2{(k_xL)}},
\end{equation}
which is an even function with respect to the incident angle, leading to 
a zero $j_y$ after the integration. However, if the effect of $\bm{m}$ is
taken into account, the quantity $|t|^2$ is no longer an even function with
respect of $\theta$, meaning that the $y$ component of the current
contributed by the electrons with incident angles $\theta$ and $-\theta$ do
not have the same magnitudes, so a net $y$ component appears. The quantity 
$|\eta_x|=|j_y/j_x|=|\sigma_{xy}/\sigma_{xx}|$ is the ratio of the Hall
conductance to the channel conductance~\cite{SDK:2014}.

Dynamical behaviors including chaos, phase synchronization, and multistability
in the coupled TI-ferromagnet system were reported in a previous 
work~\cite{WXL:2016}. Here we present a phenomenon on multistability that
was not reported in the previous work: continuous mutual switching of final
state through a sequence of multistability transitions. Specifically, we focus 
on the behavior of the system versus the bifurcation parameter $V_{dc}/V_{ac}$ 
for $\Omega/\omega_F=\Omega/(D/\hbar)=7.0$, as shown in Fig.~\ref{fig:bif}(a). 
Since $\bm{n}$ is a directional vector of unit length, it contains only 
two independent variables, e.g., $n_x$ and $n_y$, which we represent using 
the blue and red colors, respectively. We fix other parameters as 
$\xi/E=0.1$, $k_FL=100$, and $E^2eV_{ac}/(2\pi\hbar^3\omega_F\nu_F^2) = 100$. 
Figure~\ref{fig:bif}(a) shows that there are several critical parameter 
values about which the system dynamics change abruptly as reflected
by the discontinuous behaviors of the blue and red dots. A detailed 
investigation in a previous work~\cite{WXL:2016} demonstrated that 
this is a signature of multistability. 

Because the whole phase space is the surface of a 3D sphere, the relative 
strength of multistability can be characterized by the volumes of basins
of attraction of the coexisting final states (attractors). For example,
for the case of two attractors, the ratio of the volumes of their basins
of attraction indicates the relative weight of each state. 
Figures~\ref{fig:bif}(b) and \ref{fig:bif}(c) show,
for $\Omega/\omega_F = 10.0$, two representative   
basins for $V_{dc}/V_{ac}=0.5179$ and $V_{dc}/V_{ac}=0.5232$, respectively,
which are calculated by covering the unit sphere with a $100\times100$ grid
of initial conditions and determining to which attractor each initial 
condition leads to. For the two distinct attractors, the values of the  
dynamical variable $n_y$ are different: $n_y \sim 0$ and $n_y \sim -1$, 
where the sign of the driving voltage determines the sign of $n_y$. From
an applied standpoint, the two stable states are effectively binary,
which can be detected through the GMR mechanism. To label the final 
states, we color a small region on the sphere with the corresponding 
value of $n_y$ in each final state, and use Albers equal-area conic 
projection to map the sphere to a plane, which is area-preserving.
As the dc driving voltage is increased slightly, the basin of the blue 
state expands while that of the red state shrinks. As can be seen from 
the bifurcation diagram, when all the initial conditions lead to the 
blue state, we have $n_y \sim -1$. If we continue to increase the 
dc driving voltage from this point, $n_y$ will approach $0$ eventually,
indicating that the red state is the sole attractor of the system. 
During this process there is a continuous transition of multistability,
where there is a single state (blue) at the beginning, followed by 
the emergence and gradual increase of the basin of the red state, 
and finally by the disappearance of the entire basin of the blue state.
From this point on, another transition in the opposite order occurs,
where the red state eventually disappears and replaced by a blue 
state, and so on. This phenomenon of continuous flipping of the final 
state through a sequence of multistability transitions was not reported
in the previous work~\cite{WXL:2016}.   

\begin{figure}
\centering
\includegraphics[width=\linewidth]{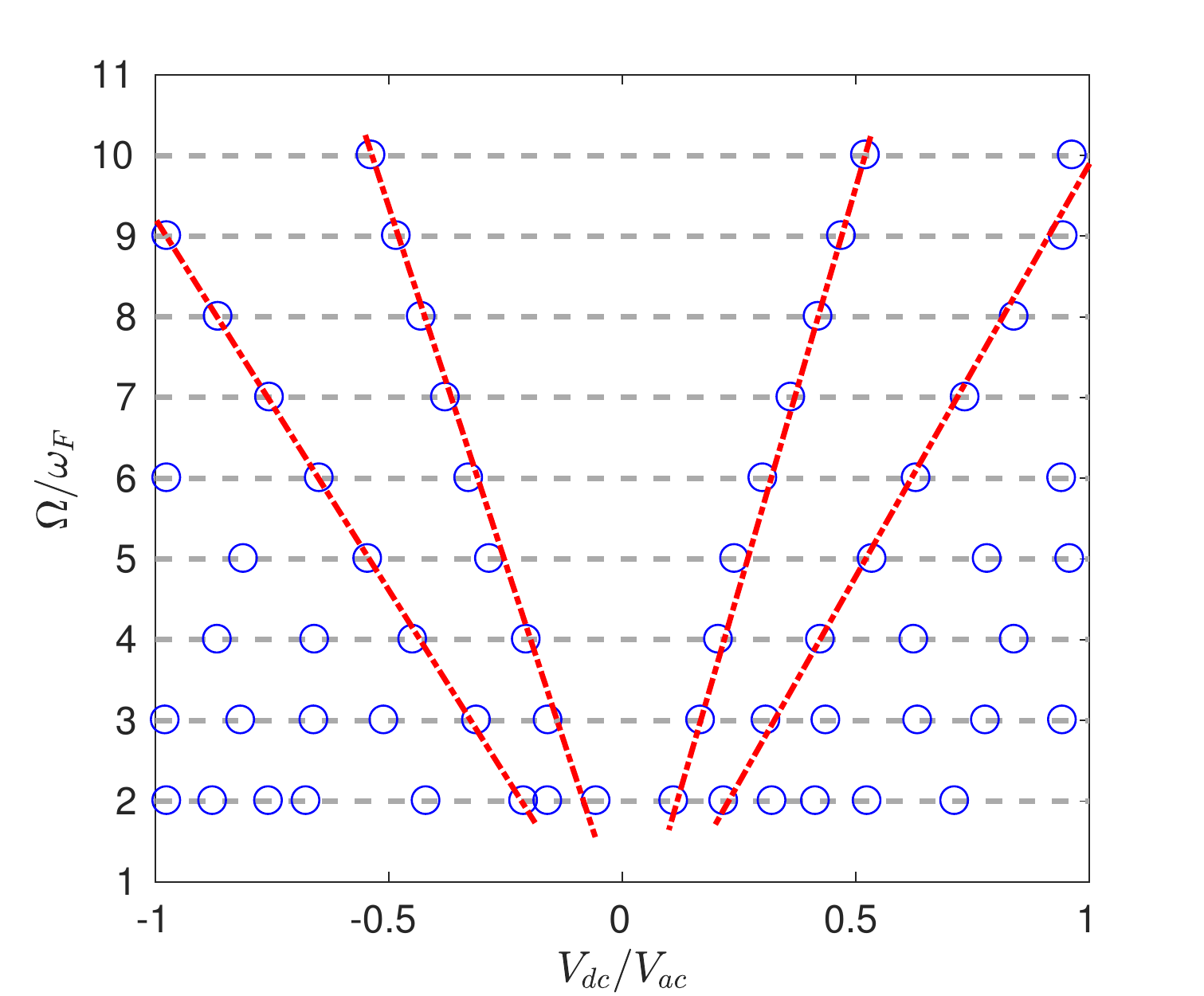}
\caption{ {\bf Dependence of multistability regimes on the driving 
voltage and frequency}. The blue circles indicate the approximate positions 
of the multistability regimes in the parameter plane of driving voltage
and frequency. The positions are those at which
discontinuous bifurcations occur with respect to the driving frequency for
the specific initial condition $\bm{n}(t=0)=(1,0,0)$. The width of each
regime depends on the condition. The red dashed lines are included for 
eye guidance. For relatively small values of $\Omega/\omega_F$ (e.g., $2.0$), 
there are more multistability regimes than those for higher values of 
$\Omega/\omega_F$. The results indicate that the emergence 
and evolution of multistability can be controlled by the driving frequency.} 
\label{fig:multi}
\end{figure}

Multistability in the coupled TI-ferromagnet system can be controlled through 
parameter perturbations. There are two approaches to altering the 
final magnetization state. We first fix a particular value of 
the driving voltage and choose the initial conditions that lead to one of the
two possible final states. Transition to the alternative final state can
be triggered by applying a perturbation, such as a voltage pulse. 
Figure~\ref{fig:multi} shows the approximate locations of the various
multistability regimes along the $V_{dc}/V_{ac}$ axis for different
values of the driving frequency. For example, for $\Omega/\omega_F = 7.0$,
there are four multistability regimes located approximately at 
the four points on the line $\Omega/\omega_F=7.0$. We then examine
the bifurcations for different values of the driving frequency and mark 
the corresponding transition points in Fig.~\ref{fig:multi}. Since the 
transition points depend on the initial condition used in calculating
the bifurcations, the results are only an approximate indicator of the 
multistability regimes of finite width. In spite of the uncertainties,
we obtain a linear behavior of the multistability regimes in the parameter 
plane of the driving frequency and voltage. We note that there are 
irregularities associated with the case of $\Omega/\omega_F=2.0$. This is 
because, in this case, the multistability regimes are too close to each other, 
rendering indistinguishable to certain extent the transition regimes.

The dependence of the multistability regimes on the driving frequency 
suggests ease to control the final state. For example, if multistability
is undesired, we can choose a relatively high value of the driving 
frequency, e.g., $\Omega/\omega_F = 10.0$. In this case, the multistability 
regimes occupy only a small part of the $V_{dc}/V_{ac}$ interval. That is,
for most parameter values there is only a single final state in the 
system, regardless of the initial conditions. A proper dc voltage can
then be chosen to guarantee that the system approaches the desired final 
state. If the task is to optimize the flexibility for the system to 
switch between distinct final states, we can set a relatively low value
of the frequency, e.g., $\Omega/\omega_F = 2.0$. In this case, the system
exhibits a large number of multistability regimes. Switching between the 
final states can be readily achieved by using a voltage pulse. 

\section{Discussion} \label{sec:discussion}

Multistability is a ubiquitous phenomenon in nonlinear 
dynamical systems~\cite{GMOY:1983,MGOY:1985,FGHY:1996,FG:1997,KFG:1999,
KF:2002,KF:2003a,KF:2003b,FG:2003,NFS:2011,Pateletal:2014,PF:2014}, and 
also in physical systems such as driven nanowire~\cite{CHLGD:2010,NYLDG:2013} 
and semiconductor superlattice~\cite{YHL:2016}. Indeed, it is common for
a nonlinear system to exhibit multiple coexisting attractors, each with its
own basin of attraction~\cite{GMOY:1983,MGOY:1985}. The boundaries among
the distinct basins can be fractal~\cite{GMOY:1983,MGOY:1985} or even 
riddled~\cite{AYYK:1992,OAKSY:1994,ABS:1994,HCP:1994,LGYV:1996,LG:1996,
ABS:1996,LA:2001,Lai:1997,BCP:1997,LG:1999b,Lai:2000}, and there is 
transient chaos~\cite{LT:book} on the basin boundaries. In applications
of nonlinear dynamics, it is thus natural to anticipate multistability.
In a specific physical system, to understand the origin of multistability 
can be beneficial to its prediction and control. Alternatively, it may be 
possible to exploit multistability for technological systems, such as 
the development of memory devices. 

The purposes of this mini-review article are twofold. First, we introduce 
topologically protected phases of matter, a frontier field in condensed 
matter physics and materials science, to the nonlinear dynamics community. 
For this purpose we provide an elementary description of a number of basic 
concepts such as Berry phase and Chern number in the context of the 
celebrated QHE with an emphasis on the topological nature, and topological 
insulators with dissipationless, spin-momentum locking surface states that 
are a remarkable source of the spin-transfer torque for nonlinear dynamical 
magnetization. As a concrete example of a hybrid topological quantum/classical 
system, we discuss the configuration of coupled TI and ferromagnet, where the 
former is a relativistic quantum system and the latter is classical. The 
physical interactions between the two types of systems are discussed: the 
spin polarized electron flows on the surface of the TI delivers a 
spin-transfer torque to the magnetization of the ferromagnet, while the 
latter modifies the Dirac Hamiltonian of the former through an exchange 
coupling. The nonlinear dynamics of this hybrid system has been studied 
previously~\cite{WXL:2016}, including a brief account of multistability. 
The second purpose is then to present results pertinent to multistability, 
which were not reported in previous works. Through a detailed parameter 
space mapping of the regions of multistability, we uncover the phenomenon 
of alternating multistability, in which the final states of the system 
emerge and disappear alternatively as some parameters are continuously 
changed. For example, by changing the frequency of the driving voltage, 
one can tune the percentage of the multistability regimes in the parameter
space. The phenomenon provides a mechanism to harness multistability through 
experimentally realizable means, such as the delivery of small voltage
pulses to the TI.

The system of coupled TI-ferromagnet is a promising prototype of the 
building blocks for the next generation of universal memory device. The 
multistable states can potentially be exploited for binary state operation
to store and process information. 

\begin{acknowledgements}

We would like to acknowledge support from the Vannevar Bush
Faculty Fellowship program sponsored by the Basic Research Office of
the Assistant Secretary of Defense for Research and Engineering and
funded by the Office of Naval Research through Grant No.~N00014-16-1-2828.

\end{acknowledgements}


\begin{thebibliography}{10}

\bibitem{Sutter:2005}
H.~Sutter, The free lunch is over: A fundamental turn toward concurrency in
  software.
\newblock {\it Dr. Dobb J.\/} {\bf 30}, 202--210 (2005).

\bibitem{CV:2015}
T.~Chouard, L.~Venema, Machine intelligence.
\newblock {\it Nature\/} {\bf 521}, 435--435 (2015).

\bibitem{LBH:2015}
Y.~LeCun, Y.~Bengio, G.~Hinton, Deep learning.
\newblock {\it Nature\/} {\bf 521}, 436--444 (2015).

\bibitem{JM:2015}
M.~I. Jordan, T.~M. Mitchell, Machine learning: Trends, perspectives, and
  prospects.
\newblock {\it Science\/} {\bf 349}, 255--260 (2015).

\bibitem{Anderson:1972}
P.~W. Anderson, More is different.
\newblock {\it Science\/} {\bf 177}, 393--396 (1972).

\bibitem{USK:2008}
K.~Uchida, M.~Saitoh, S.~Kobayashi, {\it Electron Devices Meeting, 2008. IEDM
  2008. IEEE International\/} (IEEE, 2008), pp. 1--4.

\bibitem{HYYBHYYSHF:2005}
M.~Hosomi, {\it et~al.\/}, {\it Electron Devices Meeting, 2005. IEDM Technical
  Digest. IEEE International\/} (IEEE, 2005), pp. 459--462.

\bibitem{TKMYHMOYIT:2010}
R.~Takemura, {\it et~al.\/}, A 32-mb spram with 2t1r memory cell, localized
  bi-directional write driver and1'/0'dual-array equalized reference scheme.
\newblock {\it IEEE J. Solid-State Circ.\/} {\bf 45}, 869--879 (2010).

\bibitem{BHZ:2006}
B.~A. Bernevig, T.~L. Hughes, S.-C. Zhang, Quantum spin {Hall} effect and
  topological phase transition in hgte quantum wells.
\newblock {\it Science\/} {\bf 314}, 1757--1761 (2006).

\bibitem{HK:2010}
M.~Z. Hasan, C.~L. Kane, Colloquium.
\newblock {\it Rev. Mod. Phys.\/} {\bf 82}, 3045--3067 (2010).

\bibitem{QZ:2011}
X.-L. Qi, S.-C. Zhang, Topological insulators and superconductors.
\newblock {\it Rev. Mod. Phys.\/} {\bf 83}, 1057--1110 (2011).

\bibitem{GM:2011}
P.~Gambardella, I.~M. Miron, Current-induced spin–orbit torques.
\newblock {\it Phil. Trans. R. Soc. A\/} {\bf 369}, 3175-3197 (2011).

\bibitem{MGARSPVG:2010}
I.~M. Miron, {\it et~al.\/}, Current-driven spin torque induced by the rashba
  effect in a ferromagnetic metal layer.
\newblock {\it Nat. Mater.\/} {\bf 9}, 230--234 (2010).

\bibitem{MGGZCABRSG:2011}
I.~M. Miron, {\it et~al.\/}, Perpendicular switching of a single ferromagnetic
  layer induced by in-plane current injection.
\newblock {\it Nature\/} {\bf 476}, 189--193 (2011).

\bibitem{PWBLCKS:2010}
U.~H. Pi, {\it et~al.\/}, Tilting of the spin orientation induced by rashba
  effect in ferromagnetic metal layer.
\newblock {\it Appl. Phys. Lett.\/} {\bf 97}, 162507 (2010).

\bibitem{WM:2012}
X.~Wang, A.~Manchon, Diffusive spin dynamics in ferromagnetic thin films with a
  rashba interaction.
\newblock {\it Phys. Rev. Lett.\/} {\bf 108}, 117201 (2012).

\bibitem{CMSSN:2015}
P.-H. Chang, T.~Markussen, S.~Smidstrup, K.~Stokbro, B.~K.
  Nikoli\ifmmode~\acute{c}\else \'{c}\fi{}, Nonequilibrium spin texture within
  a thin layer below the surface of current-carrying topological insulator
  ${\mathrm{bi}}_{2}{\mathrm{se}}_{3}$: A first-principles quantum transport
  study.
\newblock {\it Phys. Rev. B\/} {\bf 92}, 201406 (2015).

\bibitem{DLSK:2015}
X.~Duan, X.-L. Li, Y.~G. Semenov, K.~W. Kim, Nonlinear magnetic dynamics in a
  nanomagnet\char21{}topological insulator heterostructure.
\newblock {\it Phys. Rev. B\/} {\bf 92}, 115429 (2015).

\bibitem{TST:2015}
K.~Taguchi, K.~Shintani, Y.~Tanaka, Spin-charge transport driven by
  magnetization dynamics on the disordered surface of doped topological
  insulators.
\newblock {\it Phys. Rev. B\/} {\bf 92}, 035425 (2015).

\bibitem{JLJMLZNMSW:2015}
M.~Jamali, {\it et~al.\/}, Giant spin pumping and inverse spin {Hall} effect in
  the presence of surface and bulk spin- orbit coupling of topological
  insulator {Bi$_2$Se$_3$}.
\newblock {\it Nano Lett.\/} {\bf 15}, 7126--7132 (2015).

\bibitem{RNS:2017}
S.~Rex, F.~S. Nogueira, A.~Sudb\o{}, Topological staggered field electric
  effect with bipartite magnets.
\newblock {\it Phys. Rev. B\/} {\bf 95}, 155430 (2017).

\bibitem{LDSK:2017}
X.-L. Li, X.~Duan, Y.~G. Semenov, K.~W. Kim, Electrical switching of
  antiferromagnets via strongly spin-orbit coupled materials.
\newblock {\it J. Appl. Phys.\/} {\bf 121}, 023907 (2017).

\bibitem{Slonczewski:1996}
J.~C. Slonczewski, Current-driven excitation of magnetic multilayers.
\newblock {\it J. Magnetism Magne. Mater.\/} {\bf 159}, L1--L7 (1996).

\bibitem{MLRGMFVMKS:2014}
A.~Mellnik, {\it et~al.\/}, Spin-transfer torque generated by a topological
  insulator.
\newblock {\it Nature\/} {\bf 511}, 449--451 (2014).

\bibitem{RS:2008}
D.~C. Ralph, M.~D. Stiles, Spin transfer torques.
\newblock {\it J. Magnetism Mag. Mater.\/} {\bf 320}, 1190-1126 (2008).

\bibitem{WXL:2016}
G.-L. Wang, H.-Y. Xu, Y.-C. Lai, Nonlinear dynamics induced anomalous {Hall}
  effect in topological insulators.
\newblock {\it Sci. Rep.\/} {\bf 6}, 19803 (2016).

\bibitem{Ott:book}
E.~Ott, {\it Chaos in Dynamical Systems\/} (Cambridge University Press,
  Cambridge, UK, 2002), second edn.

\bibitem{FK:2007}
L.~Fu, C.~L. Kane, Topological insulators with inversion symmetry.
\newblock {\it Phys. Rev. B\/} {\bf 76}, 045302 (2007).

\bibitem{ZLQDFZ:2009}
H.~Zhang, {\it et~al.\/}, Topological insulators in {Bi$_2$Se$_3$, Bi$_2$Te$_3$
  and Sb$_2$Te$_3$} with a single {Dirac} cone on the surface.
\newblock {\it Nat. Phys.\/} {\bf 5}, 438--442 (2009).

\bibitem{KWBRBMQZ:2007}
M.~K{\"o}nig, {\it et~al.\/}, Quantum spin {Hall} insulator state in hgte
  quantum wells.
\newblock {\it Science\/} {\bf 318}, 766--770 (2007).

\bibitem{HQWXHCH:2008}
D.~Hsieh, {\it et~al.\/}, A topological {Dirac} insulator in a quantum spin
  {Hall} phase.
\newblock {\it Nature\/} {\bf 452}, 970--974 (2008).

\bibitem{XQHWPLBGHC:2009}
Y.~Xia, {\it et~al.\/}, Observation of a large-gap topological-insulator class
  with a single {Dirac} cone on the surface.
\newblock {\it Nat. Phys.\/} {\bf 5}, 398--402 (2009).

\bibitem{Moore:2010}
J.~E. Moore, The birth of topological insulators.
\newblock {\it Nature\/} {\bf 464}, 194--198 (2010).

\bibitem{TKNN:1982}
D.~J. Thouless, M.~Kohmoto, M.~P. Nightingale, M.~den Nijs, Quantized {Hall}
  conductance in a two-dimensional periodic potential.
\newblock {\it Phys. Rev. Lett.\/} {\bf 49}, 405--408 (1982).

\bibitem{Haldane:1988}
F.~D.~M. Haldane, Model for a quantum {Hall} effect without landau levels:
  Condensed-matter realization of the "parity anomaly".
\newblock {\it Phys. Rev. Lett.\/} {\bf 61}, 2015--2018 (1988).

\bibitem{NP:2016}
M.~Schirber, {Nobel} prize - topological phases of matter.
\newblock {\it Physics\/} {\bf 9}, 116 (2016).

\bibitem{KDP:1980}
K.~v. Klitzing, G.~Dorda, M.~Pepper, New method for high-accuracy determination
  of the fine-structure constant based on quantized {Hall} resistance.
\newblock {\it Phys. Rev. Lett.\/} {\bf 45}, 494--497 (1980).

\bibitem{PM:2012}
D.~Pesin, A.~H. MacDonald, Spintronics and pseudospintronics in graphene and
  topological insulators.
\newblock {\it Nat. Mater.\/} {\bf 11}, 409--416 (2012).

\bibitem{Hall:1879}
E.~H. Hall, On a new action of the magnet on electric currents.
\newblock {\it Ame. J. Math.\/} {\bf 2}, 287--292 (1879).

\bibitem{Laughlin:1981}
R.~B. Laughlin, Quantized {Hall} conductivity in two dimensions.
\newblock {\it Phys. Rev. B\/} {\bf 23}, 5632--5633 (1981).

\bibitem{TSG:1982}
D.~C. Tsui, H.~L. Stormer, A.~C. Gossard, Two-dimensional magnetotransport in
  the extreme quantum limit.
\newblock {\it Phys. Rev. Lett.\/} {\bf 48}, 1559--1562 (1982).

\bibitem{Laughlin:1983}
R.~B. Laughlin, Anomalous quantum {Hall} effect: An incompressible quantum
  fluid with fractionally charged excitations.
\newblock {\it Phys. Rev. Lett.\/} {\bf 50}, 1395--1398 (1983).

\bibitem{Nobel:1998}
{The Nobel Prize in Physics 1998}.

\bibitem{STG:1999}
H.~L. Stormer, D.~C. Tsui, A.~C. Gossard, The fractional quantum {Hall} effect.
\newblock {\it Rev. Mod. Phys.\/} {\bf 71}, S298--S305 (1999).

\bibitem{Hirsch:1999}
J.~E. Hirsch, Spin {Hall} effect.
\newblock {\it Phys. Rev. Lett.\/} {\bf 83}, 1834--1837 (1999).

\bibitem{MNZ:2003}
S.~Murakami, N.~Nagaosa, S.-C. Zhang, Dissipationless quantum spin current at
  room temperature.
\newblock {\it Science\/} {\bf 301}, 1348--1351 (2003).

\bibitem{SCNSJM:2004}
J.~Sinova, {\it et~al.\/}, Universal intrinsic spin {Hall} effect.
\newblock {\it Phys. Rev. Lett.\/} {\bf 92}, 126603 (2004).

\bibitem{BZ:2006}
B.~A. Bernevig, S.-C. Zhang, Quantum spin {Hall} effect.
\newblock {\it Phys. Rev. Lett.\/} {\bf 96}, 106802 (2006).

\bibitem{Datta:book}
S.~Datta, {\it Electronic Transport in Mesoscopic Systems\/} (Cambridge
  University Press, Cambridge, England, 1995).

\bibitem{KBWHLQZ:2008}
M.~K{\"o}nig, {\it et~al.\/}, The quantum spin {Hall} effect: theory and
  experiment.
\newblock {\it J. Phys. Soc. Japan\/} {\bf 77}, 031007 (2008).

\bibitem{LCG:2014}
N.~Locatelli, V.~Cros, J.~Grollier, Spin-torque building blocks.
\newblock {\it Nat. Mater.\/} {\bf 13}, 11--20 (2014).

\bibitem{GNSTOAGHBZ:2012}
M.~Gajek, {\it et~al.\/}, Spin torque switching of 20 nm magnetic tunnel
  junctions with perpendicular anisotropy.
\newblock {\it Appl. Phys. Lett.\/} {\bf 100}, 132408 (2012).

\bibitem{BKO:2012}
A.~Brataas, A.~D. Kent, H.~Ohno, Current-induced torques in magnetic materials.
\newblock {\it Nat. Mater.\/} {\bf 11}, 372--381 (2012).

\bibitem{BBFVPECFC:1988}
M.~N. Baibich, {\it et~al.\/}, Giant magnetoresistance of (001)fe/(001)cr
  magnetic superlattices.
\newblock {\it Phys. Rev. Lett.\/} {\bf 61}, 2472--2475 (1988).

\bibitem{BGSZ:1989}
G.~Binasch, P.~Gr\"unberg, F.~Saurenbach, W.~Zinn, Enhanced magnetoresistance
  in layered magnetic structures with antiferromagnetic interlayer exchange.
\newblock {\it Phys. Rev. B\/} {\bf 39}, 4828--4830 (1989).

\bibitem{Nobel:2007}
{The Nobel Prize in Physics 2007}.

\bibitem{MGGZC:2011}
I.~M. Miron, {\it et~al.\/}, Perpendicular switching of a single ferromagnetic
  layer induced by in-plane current injection.
\newblock {\it Nature\/} {\bf 476}, 189--193 (2011).

\bibitem{LPLTRB:2012}
L.~Liu, {\it et~al.\/}, Spin-torque switching with the giant spin {Hall} effect
  of tantalum.
\newblock {\it Science\/} {\bf 336}, 555--558 (2012).

\bibitem{SJLBAPBMG:2016}
C.~Safeer, {\it et~al.\/}, Spin--orbit torque magnetization switching
  controlled by geometry.
\newblock {\it Nat. Nanotech.\/} {\bf 11}, 143--146 (2016).

\bibitem{Yokoyama:2011}
T.~Yokoyama, Current-induced magnetization reversal on the surface of a
  topological insulator.
\newblock {\it Phys. Rev. B\/} {\bf 84}, 113407 (2011).

\bibitem{SDK:2014}
Y.~G. Semenov, X.~Duan, K.~W. Kim, Voltage-driven magnetic bifurcations in
  nanomagnet--topological insulator heterostructures.
\newblock {\it Phys. Rev. B\/} {\bf 89}, 201405 (2014).

\bibitem{GMOY:1983}
C.~Grebogi, S.~W. McDonald, E.~Ott, J.~A. Yorke, Final state sensitivity: an
  obstruction to predictability.
\newblock {\it Phys. Lett. A\/} {\bf 99}, 415-418 (1983).

\bibitem{MGOY:1985}
S.~W. McDonald, C.~Grebogi, E.~Ott, J.~A. Yorke, Fractal basin boundaries.
\newblock {\it Physica D\/} {\bf 17}, 125-153 (1985).

\bibitem{FGHY:1996}
U.~Feudel, C.~Grebogi, B.~R. Hunt, J.~A. Yorke, Map with more than 100
  coexisting low-period periodic attractors.
\newblock {\it Phys. Rev. E\/} {\bf 54}, 71--81 (1996).

\bibitem{FG:1997}
U.~Feudel, C.~Grebogi, Multistability and the control of complexity.
\newblock {\it Chaos\/} {\bf 7}, 597-604 (1997).

\bibitem{KFG:1999}
S.~Kraut, U.~Feudel, C.~Grebogi, Preference of attractors in noisy multistable
  systems.
\newblock {\it Phys. Rev. E\/} {\bf 59}, 5253--5260 (1999).

\bibitem{KF:2002}
S.~Kraut, U.~Feudel, Multistability, noise, and attractor hopping: The crucial
  role of chaotic saddles.
\newblock {\it Phys. Rev. E\/} {\bf 66}, 015207 (2002).

\bibitem{KF:2003a}
S.~Kraut, U.~Feudel, Enhancement of noise-induced escape through the existence
  of a chaotic saddle.
\newblock {\it Phys. Rev. E\/} {\bf 67}, 015204(R) (2003).

\bibitem{KF:2003b}
S.~Kraut, U.~Feudel, Noise-induced escape through a chaotic saddle: Lowering of
  the activation energy.
\newblock {\it Physica D\/} {\bf 181}, 222-234 (2003).

\bibitem{FG:2003}
U.~Feudel, C.~Grebogi, Why are chaotic attractors rare in multistable systems?
\newblock {\it Phys. Rev. Lett.\/} {\bf 91}, 134102 (2003).

\bibitem{NFS:2011}
C.~N. Ngonghala, U.~Feudel, K.~Showalter, Extreme multistability in a chemical
  model system.
\newblock {\it Phys. Rev. E\/} {\bf 83}, 056206 (2011).

\bibitem{Pateletal:2014}
M.~S. Patel, {\it et~al.\/}, Experimental observation of extreme multistability
  in an electronic system of two coupled {R}\"ossler oscillators.
\newblock {\it Phys. Rev. E\/} {\bf 89}, 022918 (2014).

\bibitem{PF:2014}
A.~N. Pisarchik, U.~Feudel, Control of multistability.
\newblock {\it Phys. Rep.\/} {\bf 540}, 167-218 (2014).

\bibitem{CHLGD:2010}
Q.~Chen, L.~Huang, Y.-C. Lai, C.~Grebogi, D.~Dietz, Extensively chaotic motion
  in electrostatically driven nanowires and applications.
\newblock {\it Nano lett.\/} {\bf 10}, 406--413 (2010).

\bibitem{NYLDG:2013}
X.~Ni, L.~Ying, Y.-C. Lai, Y.~Do, C.~Grebogi, Complex dynamics in nanosystems.
\newblock {\it Phys. Rev. E\/} {\bf 87}, 052911 (2013).

\bibitem{YHL:2016}
L.~Ying, D.~Huang, Y.-C. Lai, Multistability, chaos, and random signal
  generation in semiconductor superlattices.
\newblock {\it Phys. Rev. E\/} {\bf 93}, 062204 (2016).

\bibitem{AYYK:1992}
J.~C. Alexander, J.~A. Yorke, Z.~You, I.~Kan, Riddled basins.
\newblock {\it Int. J. Bifur. Chaos Appl. Sci. Eng.\/} {\bf 2}, 795-813 (1992).

\bibitem{OAKSY:1994}
E.~Ott, J.~C. Alexander, I.~Kan, J.~C. Sommerer, J.~A. Yorke, The transition to
  chaotic attractors with riddled basins.
\newblock {\it Physica D\/} {\bf 76}, 384-410 (1994).

\bibitem{ABS:1994}
P.~Ashwin, J.~Buescu, I.~Stewart, Bubbling of attractors and synchronisation of
  oscillators.
\newblock {\it Phys. Lett. A\/} {\bf 193}, 126-139 (1994).

\bibitem{HCP:1994}
J.~F. Heagy, T.~L. Carroll, L.~M. Pecora, Experimental and numerical evidence
  for riddled basins in coupled chaotic systems.
\newblock {\it Phys. Rev. Lett.\/} {\bf 73}, 3528-3531 (1994).

\bibitem{LGYV:1996}
Y.-C. Lai, C.~Grebogi, J.~A. Yorke, S.~Venkataramani, Riddling bifurcation in
  chaotic dynamical systems.
\newblock {\it Phys. Rev. Lett.\/} {\bf 77}, 55-58 (1996).

\bibitem{LG:1996}
Y.-C. Lai, C.~Grebogi, Noise-induced riddling in chaotic dynamical systems.
\newblock {\it Phys. Rev. Lett.\/} {\bf 77}, 5047-5050 (1996).

\bibitem{ABS:1996}
P.~Ashwin, J.~Buescu, I.~Stewart, From attractor to chaotic saddle: a tale of
  transverse instability.
\newblock {\it Nonlinearity\/} {\bf 9}, 703-737 (1996).

\bibitem{LA:2001}
Y.-C. Lai, V.~Andrade, Catastrophic bifurcation from riddled to fractal basins.
\newblock {\it Phys. Rev. E\/} {\bf 64}, 056228 (2001).

\bibitem{Lai:1997}
Y.-C. Lai, Scaling laws for noise-induced temporal riddling in chaotic systems.
\newblock {\it Phys. Rev. E\/} {\bf 56}, 3897-3908 (1997).

\bibitem{BCP:1997}
L.~Billings, J.~H. Curry, E.~Phipps, Lyapunov exponents, singularities, and a
  riddling bifurcation.
\newblock {\it Phys. Rev. Lett.\/} {\bf 79}, 1018-1021 (1997).

\bibitem{LG:1999b}
Y.-C. Lai, C.~Grebogi, Riddling of chaotic sets in periodic windows.
\newblock {\it Phys. Rev. Lett.\/} {\bf 83}, 2926-2929 (1999).

\bibitem{Lai:2000}
Y.-C. Lai, Catastrophe of riddling.
\newblock {\it Phys. Rev. E\/} {\bf 62}, R4505--R4508 (2000).

\bibitem{LT:book}
Y.-C. Lai, T.~T\'{e}l, {\it Transient Chaos - Complex Dynamics on Finite Time
  Scales\/} (Springer, New York, 2011).

\end{thebibliography}

\end{document}